\documentclass[seceq,preprint]{ptptex}

\usepackage{latexsym}

\title{
Notes on the Super Nambu Bracket
}

\author{
Masato {\sc Sakakibara}
\footnote{E-mail: sakakiba@monet.phys.s.u-tokyo.ac.jp}
}

\inst{
Department of Physics, University of Tokyo, 
Tokyo 113-0033, Japan}

\recdate{
September 4, 2002}

\abst{
We define a super Nambu-Poisson algebra over a super manifold. 
A super Nambu bracket does not satisfy the usual skew-symmetric property, 
and we propose another skew-symmetric property.   
We show that the divergence of super Nambu-Hamiltonian vector fields leads 
to a generalization of the Batalin-Vilkovisky algebra. 
}

\notypesetlogo

\def\a{\alpha}

\def\i{\begin{eqnarray}}
\def\f{\end{eqnarray}}
\def\non{\nonumber}
\def\del#1#2{\frac{\partial{#1}}{\partial {#2}}}
\def\delr#1#2{\frac{\partial_{r}{#1}}{\partial {#2}}}
\def\dell#1#2{\frac{\partial_{l}{#1}}{\partial {#2}}}

\def\d{\partial}

\def\R{{\Bbb R}}
\def\Z{{\Bbb Z}}
\def\D{\Delta}

\def\ep{\epsilon}
\def\t{\tau}
\def\th{\theta}
\def\w{\omega}
\def\g{\frak{g}}
\def\vs{\vspace}
\def\lra{\longrightarrow}
\def\no{\noindent}
\def\we{\wedge}
\def\s{\sigma}
\def\hs{\hspace}
\def\CM{C^{\infty}(M)}

\def\theequ
ation{\thesection.\arabic{equation}}
\makeatletter
\@addtoreset{equation}{section}
\makeatother 

\begin{document}

\maketitle

\section{Introduction}

The Nambu-Hamiltonian (NH) system is a generalization of the Hamilton system 
that was introduced by Nambu \cite{Nam1}.
Many authors\cite{Tak1} have studied the fundamental properties of 
the NH system of bosons.  
In order to add fermions to the system, we must 
extend the Nambu bracket to a super Nambu bracket (SNB). 
A SNB does not satisfy the usual skew-symmetric 
property, and we propose another skew-symmetric property.
In this paper, we demonstrate the three properties that a SNB satisfies, and 
define a super Nambu-Poisson algebra on a super manifold. 
Some authors have remarked on the relations\cite{Tak1,Sta1} 
between the Nambu-Poisson algebra and the $L_{\infty}$ algebra, or the
Batalin-Vilkovisky (BV) 
algebra,\cite{BV1} especially with regard to the Nambu bracket and 
higher brackets of the $L_{\infty}$ algebra.
In order to determine these relations, 
we extend the Nambu-Poisson algebra to a $\Z_2$-graded algebra.
We show that the divergence of the NH vector fields of a SNB
leads to a generalization of the BV algebra.
Throughout this paper, 
$\d_l/\d x$ (resp.,$\;\d_r/\d x$) denotes a left (resp.,$\;$right) 
derivative, and $|f|=0,1$ is the degree (Grassmann parity) of $f$, modulo $2$.

\section{Super Nambu-Poisson algebra}
An example of the SNB over $\R^{2|1}$ with coordinates $(x_1,x_2,\th)$ 
is  
\i\label{2:0} \{f,g,h\}&=& 
(-)^{|g|}\left(\del{f}{x_1}\delr{g}{\th}\del{h}{x_2}
-\del{f}{x_2}\delr{g}{\th}\del{h}{x_1}\right)+\left(\delr{f}{\th}
\del{g}{x_1}\del{h}{x_2}\right.\non\\
& &\hs{-2cm}-\left.\delr{f}{\th}\del{g}{x_2}\del{h}{x_1}\right)
+(-)^{|g|+|h|}\left(\del{f}{x_2}\del{g}{x_1}\delr{h}{\th} 
-\del{f}{x_1}\del{g}{x_2}\delr{h}{\th}\right), \f
with degree $\ep=1$.
This bracket satisfies the $\Z_2$-graded Nambu-Poisson algebra.
In this section, we define the even and odd super Nambu-Poisson algebra.

\subsection{Super Nambu-Poisson manifold}

The bosonic Nambu bracket is characterized by the following
three properties:\cite{Tak1}
 (i) skew-symmetry, (ii) the Leibniz rule, and 
(iii) the fundamental identity (FI). 
A manifold with a Nambu bracket is called a Nambu-Poisson manifold.
We define a SNB $\{\cdot,\cdots,\cdot\}$ as a bracket over 
a $d$-dimensional super manifold $M$
that satisfies three properties with the $\Z_2$-grading. 
First, we assume there exist SNBs and demonstrate these properties as the 
necessary conditions for this existence. 
We define a Nambu-Poisson tensor\footnote{We do not assume that 
$\eta$ is an element of $\Lambda^{n} TM$.}
$\eta\in TM^{\otimes n}$ 
as $\eta(df_1,\cdots,df_n)=\{f_1,\cdots,f_n\}$ for any $f_i\in \CM$ 
and the map $\sharp:T^{\ast}M^{\otimes n-1}\lra TM$ as
\i \sharp(\a_1\otimes\cdots\otimes \a_{n-1})=\eta(\,\cdot\,,\a_1,\cdots,\a_{n-1}), \f
for $\a_i\in T^{\ast}M$.
Owing to the action principle\cite{Tak1} of the NH system,  
we assume that a NH vector field is given by 
\i X_{f_1\cdots f_{n-1}}=\sharp(df_1\otimes\cdots\otimes df_{n-1})=
\frac{1}{(n-1)!}\,\sharp(df_1\we\cdots\we df_{n-1}).\f
The degree of a NH field is 
\i \left|X_{f_1\cdots f_{n-1}}\right|=\ep+\sum_{i=1}^{n-1}|f_i|, \f 
 where $\ep=|\eta|=0,1$, and the degree of a SNB is 
\i \left|\{f_1,\cdots,f_n\}\right|=\left|df_1(X_{f_2\cdots f_n})\right|=
\ep+\sum_{i=1}^{n}|f_i|. \f
Below we demonstrate three properties that a SNB satisfies under these assumptions.

\vs{3mm}

\no
{\bf (i) The skew-symmetric property} 

\vs{2mm}

\no
Owing to the skew-symmetry of $df_i\we df_{i+1}$, 
a SNB possesses the skew-symmetric property 
\i\label{2:6} 
\{f_1,\cdots,f_i,f_{i+1},\cdots\;\}=
-(-)^{|f_i||f_{i+1}|}\{f_1,\cdots,f_{i+1},f_i,\cdots\;\}, \f
for $i=2,\cdots,n-1$. 
Note that we cannot impose the same condition for $i=1$ and $\ep=1$.
For $n=2$ and $i=1$, this property differs from the skew-symmetry 
of an odd Poisson bracket.

\vs{2mm}

\no
{\bf Remark} For a class of bosonic Nambu brackets $\{f_1,f_2,f_3\}$
over a ``symplectic'' 
manifold $(M,\w^{(3)})$,  
where $\w^{(3)}$ is a non-degenerate\footnote 
{If the equation $i(X)\w^{(3)}=df_1\we df_2$, where $i$ denotes 
an interior product, has a solution $X$, we 
say that $\w^{(3)}$ is non-degenerate.} 
closed $3$-form, the Nambu bracket also possesses the following skew-symmetric property. 
A usual Poisson bracket has the skew-symmetric property, owing to the equation 
$\{f,g\}=\w(X_{f},X_{g})$.
For a bosonic Nambu bracket of this class, we have the equation 
\i\label{2:10} 
\w^{(3)}(X_{f_1 f_2},X_{g_1 g_2},X_{h_1 h_2})=
(df_1\we df_2) (X_{g_1,g_2},X_{h_1,h_2})\non\\
&&\hs{-5cm}=\{f_1,g_1,g_2\}\{f_2,h_1,h_2\}-
\{f_2,g_1,g_2\}\{f_1,h_1,h_2\}, \f
for any $f_i,g_i$ and $h_i$ that do not depend on 
the local coordinates.
Since the LHS of (\ref{2:10}) is skew-symmetric, we have 
\i \w^{(3)}(X_{f_1 f_2},X_{g_1 g_2},X_{h_1 h_2})=\;
 -\w^{(3)}(X_{g_1 g_2},X_{f_1 f_2},X_{h_1 h_2}).  \f
The RHS of (\ref{2:10}) has the same skew-symmetric property,
\i\label{2:12}
&&\{f_1,g_1,g_2\}\{f_2,h_1,h_2\}-\{f_2,g_1,g_2\}\{f_1,h_1,h_2\} \non\\
&&\hs{3cm}=-\{g_1,f_1,f_2\}\{g_2,h_1,h_2\}+\{g_2,f_1,f_2\}\{g_1,h_1,h_2\},
\f
and so on. 
We can easily extend the property (\ref{2:12})
to the case that $n>3$ and to a SNB. 
This fact implies
that we should impose the above property instead of the usual skew-symmetric property
on the general (bosonic and super) Nambu bracket.

\vs{2mm}

\no
{\bf (ii) The Leibniz rule} 

\vs{2mm}

\no
Owing to the Leibniz rule for $X_{f_1,\cdots,f_{n-1}}$,
a SNB satisfies the equation,
\i \label{2:3} 
&&\hs{-2cm}\{gh,f_2,\cdots,f_n\} \non\\
&=& g\{h,f_2,\cdots,f_n\}+
(-)^{(\ep+\sum_{i=2}^n|f_i|)|h|}\{g,f_2,\cdots,f_n\}h. \f
For $i\ge 2$, from the Leibniz rule for an exterior derivative and the linearity of the map $\sharp$, 
we obtain the Leibniz rule 
\i \label{2:4} 
&&\hs{-2cm}\{f_1,f_2,\cdots,\stackrel{i}{\check{gh}},
\cdots,f_{n-1}\} 
= (-)^{(\sum_{j=i+1}^{n-1}|f_j|)|h|} \{f_1,f_2,\cdots,g,\cdots,f_n\}h \non\\
&+& (-)^{(|h|+\sum_{j=i+1}^n|f_j|)|g|}\{f_1,f_2,\cdots,h,\cdots,f_n\}g. \f
Note that this Leibniz rule leads to a $\Z_2$-graded 
cyclic cocycle-{\it like} property of a SNB,
\i\label{2:9} 
\sum_{i=1}^{n-1}(-)^{i+1}\{f,f_1,&\cdots&,f_i 
f_{i+1},\cdots,f_n\} \non \\
&+&(-)^{n+|f_n|(\sum_{j=1}^{n-1}|f_j|)}\{f,f_n f_1,\cdots,f_{n-1}\}=0, \f
for any fixed $f$. This property is a necessary condition for (\ref{2:4}) to hold.  

\vs{2mm}

\no
{\bf (iii) The fundamental identity}

\vs{2mm}

\no
The FI is generalized as 
\i\label{2:5} 
&&\hs{-1.5cm}\{\{g_1,\cdots,g_n\},f_1,\cdots,f_{n-1}\} \non\\
&=&\sum_{i=2}^{n-1}(-)^{(\ep+\sum_{j=1}^{n-1}|f_j|)(\sum_{k=i+1}^n|g_k|)}
\{g_1,\cdots,\{g_i,f_1,\cdots,f_{n-1}\},\cdots,g_n\} \non\\
&+&(-)^{(\ep+\sum_{j=1}^{n-1}|f_j|)(\ep+\sum_{k=2}^n|g_k|)}
\{\{g_1,f_1,\cdots,f_{n-1}\},g_2,\cdots,g_n\}.\f
This property implies that the SNB satisfies the Leibniz rule for the SNB itself. 
\vs{2mm}

Here, we obtain the three properties of the SNB. We also 
define the SNB as the bracket that satisfies 
all three of these properties and define the super Nambu-Poisson manifold as the 
manifold that is equipped with a SNB.

\vspace{3mm}

\no
{\bf Definition} A super Nambu-Poisson manifold is a super manifold $M$ 
with a polylinear map, called a super Nambu bracket,
$\{,\cdots,\}:A^{\otimes n}\lra A$, where $A$ is $C^{\infty}(M)$, 
that satisfies (i) the skew-symmetric property given in (\ref{2:6}),   
(ii) the Leibniz rule given in (\ref{2:3}) and 
(\ref{2:4}), and (iii) the fundamental identity given in (\ref{2:5}).
We call the algebra $(A,\{,\cdots,\})$ a super Nambu-Poisson algebra. 

\vspace{3mm}

\no
An example of a super Nambu-Poisson algebra is the algebra with the odd SNB (\ref{2:0}) over $\R^{2|1}$ 
that satisfies the above three properties and (\ref{2:12}) with the $\Z_2$-grading.
Note that the super Jacobian in this case is not a SNB, unlike in the bosonic case. 
This results from the difference between the differential forms and 
the volume forms over a super manifold.

\vspace{3mm}

\no
{\bf Remarks}
(i) A SNB satisfies all of our assumptions. 
(ii) The order $n$ of a SNB need not satisfy $n\le \dim M$.  
(iii) For a bosonic Nambu bracket, the $n-p$ order bracket defined by 
$\{f_1,\cdots,f_p\}':=
\{f_1,\cdots,f_p,g_1,\cdots,g_{n-p} \}$ for fixed $\{g_i\}$ is also a
Nambu bracket. For a SNB, this is not true, because a bracket $\{,\cdots,\}'$ 
does not satisfy the FI, as $\{\cdots,f,f,\cdots\}\neq 0$ in general.

\subsection{Decomposition of the super Nambu bracket}

The fact that SNBs do not possess the usual skew-symmetric property  
suggests a decomposition of the SNB, as in the case of a bosonic canonical Nambu bracket.\cite{Sak1} 
In fact, Theorem 1 in Ref.~\citen{Sak1} can be generalized to the SNB as follows.

\vspace{3mm}

\no
{\bf Theorem 1} Let $\g$ be a super Lie algebra with degree $\ep$.
Let $\t$ be a degree 0 polylinear map $\g^{\otimes n-1}
\lra\g$ that is 
skew-symmetric: 
\i \t(\cdots,a,b,\cdots)=-(-)^{|a||b|}\t(\cdots,b,a,\cdots). \f
We assume that this map satisfies the equation
\i\label{2:8} [a,\t(b_1,\cdots,b_{n-1})]=(-)^{(|a|+\ep)
\sum_{j=i+1}^{n-1}|b_j|}
\sum _{i=1}^{n-1}\t(b_1,\cdots,[a,b_i],\cdots,b_{n-1}). \f
Then, the degree $\ep$ bracket $\{a_1,a_2\cdots,a_n\}
:=[a_1,\t(a_2,\cdots,a_n)]$ 
is a polylinear map $\g^{\otimes n}\lra\g$ that satisfies 
the skew-symmetry (\ref{2:6}) and the fundamental identity (\ref{2:5}). 

\vspace{2mm}

\no
We can easily prove this theorem using (\ref{2:8}) and the 
Jacobi identity of $\g$.
Note that the set of  all linear combinations of elements 
$\t(a_1,\cdots,a_{n-1})$ is a Lie sub-algebra of $\g$.

\vspace{2mm}

\no
{\bf Remarks} (i) The bracket $\{,\cdots,\}$ is not skew-symmetric in 
the sense of (\ref{2:12}).
(ii) Equation (\ref{2:9}) and the Leibniz rule (\ref{2:3})
suggest that we extend $\g$ to a Poisson algebra and impose the equation
\i\label{cyclic2} \sum_{i=1}^{n-1}(-)^{i+1}\t(a_1,&\cdots&,a_i a_{i+1},
\cdots,a_n) \non\\
&+&(-)^{n+|a_n|(\sum_{i=1}^{n-1}|a_i|)}\t(a_n a_1,\cdots,a_{n-1})=0.\f
In fact, for the bosonic canonical Nambu bracket,\cite{Sak1} the Lie bracket 
corresponds to the Poisson (Dirac) bracket, and
the map $\t$ corresponds to the cyclic cocycle over the algebra $C^{\infty}(T^{n-2})$.

\section{Generalization of the Batalin-Vilkovisky Algebra}

Over an 
odd symplectic manifold $M$ with the coordinates $(z_1,\cdots,z_{d})$, 
the Hamilton vector fields need not be divergenceless, 
and the divergence plays an important role in the geometric realization of 
the BV quantization.\cite{BV1} The volume element $\mu$ on $M$ is specified by 
the density function $\rho(z)$. The divergence of the Hamilton vector field 
$X_{H}=\delr{}{z_i}X_H^i$ is determined by\cite{Sch1}   
\i\label{3:1}
{\rm div}_{\mu}X_H=2\Delta H=\sum_i(-)^{|z_i|}\rho^{-1}
\dell{}{z_i}(\rho X^i_H), \f
and the anti-bracket satisfies\cite{Kos1,Wit1}
\i\label{3:2} 
(-)^{|f|}\{f,g\}=\D(fg)-\D(f)g-(-)^{|f|}f\D(g).\f 
The divergence $\D$ satisfies the Leibniz rule for the anti-bracket: 
\i\label{deriv} \D(\{f,g\})=\{\D(f),g\}+(-)^{|f|+1}\{f,\D(g)\}. \f
In the BV quantization, we impose the nilpotency condition $\D^2=0$, 
which gives the condition on $\rho$.

On the super Nambu-Poisson manifold with the volume element $\mu$, 
the NH vector fields need not be divergenceless  
over both an even and an odd Nambu-Poisson manifold.
We can define the divergence of a NH vector field in the same same way as (\ref{3:1}):
\i {\rm div}_{\mu} X_{H_1,\cdots,H_{n-1}}=:2\Delta(H_1,\cdots,H_{n-1}). \f
This divergence has the skew-symmetric property
\i \Delta(\cdots,f,g,\cdots)=-(-)^{|f||g|}\Delta(\cdots,g,f,\cdots) \f
and its degree is $\ep$. 
For example, we consider an even SNB over $\R^{1|2}$ with coordinates 
$(x,\th_1,\th_2)$, 
\i && \{f,g,h\}=(-)^{1+|g|}\del{f}{x}\left(\delr{g}{\th_1}
\delr{h}{\th_2}+\delr{g}{\th_2}\delr{h}{\th_1}\right)
+(-)^{|g|+|h|}\delr{f}{\th_1}\left(\del{g}{x}\delr{h}{\th_2}\right. \non\\
&&\hs{2cm}\left.-(-)^{|h|}\delr{g}{\th_2}\del{h}{x}\right)
+(-)^{|g|+|h|}\delr{f}{\th_2}\left(\del{g}{x}\delr{h}{\th_1}-(-)^{|h|}
\delr{g}{\th_1}\del{h}{x}\right). \f
In this case, the divergence is given by 
\i\label{div} \Delta(f,g)&=&(-)^{|g|+1}\left(
\frac{\d_r^2 f}{\d x\d \th_1}\delr{g}{\th_2}
+\delr{f}{\th_j}\frac{\d_r^2 g}{\d x\d\th_2} \right. \non\\ 
&&\hs{2cm}+\left.\frac{\d_r^2f}{\d x\d \th_2}\delr{g}{\th_1}+ 
\delr{f}{\th_2}\frac{\d_r^2 g}{\d x\d\th_1}\right).\f

Here, we give the properties of the divergence. 
By the Leibniz rule (\ref{2:4}), the NH vector fields satisfy
\i\label{3:3} 
X_{fg,f_2,\cdots,f_{n-1}}&=&(-)^{|g||F|}X_{f,f_2,\cdots,f_{n-1}}g
+(-)^{|f|(|g|+|F|)} X_{g,f_2,\cdots,f_{n-1}}f, \f
where $|F|=\sum_{i=2}^{n-1}|f_i|$, and $Xf$ represents $(Xf)(g)=(dg)(X)\cdot f$. 
Taking the divergence of both sides of (\ref{3:3}), 
and using the equation $\sum_i\delr{f_1}{z_i} X_{f_2,\cdots,f_n}^i
=\{f_1,\cdots,f_n\}$, we obtain
\i\label{3:4} 
&&\hs{-2cm}\frac{1}{2}\left((-)^{|f|\ep}\{f,g,f_2,\cdots,f_{n-1}\}
-(-)^{(|f|+\ep)|g|+1}\{g,f,f_2,\cdots,f_{n-1}\}\right)\non\\
&=&\Delta(fg,f_2,\cdots,f_{n-1})-(-)^{|g||F|}\Delta(f,f_2,\cdots,f_{n-1})
g\non\\
& &\hs{2cm}-(-)^{|f|\ep}\,f\Delta(g,f_2,f_3,\cdots,f_{n-1}).\f
For $n=2$, the above equation is reduced to (\ref{3:2}). 
(A similar equation\footnote{By the skew-symmetric property of $\D$ and the 
identification ($C^{\infty}(M),\ep,\D$) with (${\cal A},|\D|,\Phi^{n-1}_{\D}$) given 
in Ref.~\citen{Akm1},
the RHS of (\ref{3:4}) is shown to be the same as that of (7) in Ref.~\citen{Akm1}, 
up to a constant factor.} appears in Refs.~\citen{Kos1} and~\citen{Akm1}.)
In contrast to the case of the anti-bracket, in the present case we cannot define the SNB in terms of 
$\D$, because the SNB does not have the usual skew-symmetric property.
By the FI (\ref{2:5}), the NH vector field satisfies
\i\label{3:5}
&&[X_{g_1, \cdots,g_{n-1}},X_{f_1,\cdots,f_{n-1}}] \non\\
&&\hs{1cm}=\sum_{i=1}^{n-1}(-)^{(\ep+|F|)(\sum_{k=i+1}^{n-1}|g_k|)+1}
X_{g_1,\cdots,\{g_i,f_1,\cdots,f_{n-1}\},\cdots,g_{n-1}}, \f
where $|F|=\sum_{i=1}^{n-1}|f_i|$, and $XY$ represents $XY(f)= d(df(X))(Y)$. 
Taking the divergence of both sides of (\ref{3:5}), we obtain
\i&&\hs{-1.5cm}\{\D(f_1,\cdots,f_{n-1}),g_1,\cdots,g_{n-1}\}\non\\
&=&\sum_{i=1}^{n-1} (-)^{(\ep+|F|)(\ep+\sum_{j=i+1}^{n-1}|g_j|)+1} 
\D(g_1,\cdots,\{g_i,f_1,\cdots,f_{n-1}\},
\cdots,g_{n-1})\non\\
& &\hs{2cm}+(-)^{(\ep+|F|)(\ep+|G|)}\{\D(g_1,\cdots,g_{n-1}),f_1,
\cdots,f_{n-1}\},\f
where $|G|=\sum_{i=1}^{n-1}|g_i|$.
For $n=2$, the above equation reduces to (\ref{deriv}). 
We want to generalize the nilpotency condition for $\D$.
However, for example, 
the divergence $(\ref{div})$ is not nilpotent as a bi-differential operator nor as a co-derivation.
Note that if the SNB is given by the Lie (or Poisson) bracket and the map 
$\t$ of Theorem 1, 
we obtain the relations between them and $\D$. 
This generalized BV algebra should be useful for 
the quantization of the NH system using the BV method, and extended objects.

\section{Discussion}

In this paper, we defined the super Nambu-Poisson algebra and demonstrated
its connection with the generalized BV algebra.
Here, we comment on some physical applications. 
The super Nambu-Hamilton system has the equation of motion 
\i \frac{df}{dt}=\{f,h_1,\cdots,h_{n-1}\}, \f
with the Hamiltonians $h_1,\cdots,h_{n-1}$. 

One trivial example is a system  with one free boson $(x,p)\!=\!(x_1,x_2)$ 
and one free fermion $\th$ over 
the phase space $\R^{2|1}$.
In this case, the equation of motion is given by the SNB (\ref{2:0}) and the Hamiltonians 
$h_1=\frac{1}{2}p^2,\; h_2=\th$.
Other non-trivial examples, including Euler's top\cite{Nam1}, 
have not yet been elucidated and must be studied. 
We consider a Lagrangian system\cite{Hop1}, such as that with the Polyakov action 
with the light-cone gauge,
that is invariant under the  $n$-dimensional volume preserving diffeomorphism
\i \delta_{\ep} X_{\mu}(\s)=\{X_{\mu},\ep_1,\cdots,\ep_{n-1}\}, \f
where the bracket is the bosonic Nambu bracket. 
To apply the BRS or BV method to such a constrained Lagrangian system,  
the parameters $\{\ep_j\}$ must be replaced by the ``ghosts" $\{c_j\}$, which 
are fermions, and the above Nambu bracket becomes the SNB with super coordinates.  
It would be interesting to study the supersymmetric extension of the above 
Lagrangian with the SNB.
The quantization of the SNB involves the same difficulties as that of the bosonic Nambu bracket.
\cite{Nam1,Tak1} 
The triple brackets\cite{Nam1} are meaningless for our Nambu bracket, 
which does not possess the usual skew-symmetric property. 
The difficulty\cite{Nam1} that the triple bracket 
for the bosonic Nambu bracket does 
not satisfy the Leibniz rule may arise from this fact.
On the other hand, to quantize the Nambu bracket, 
we should construct the representation of the algebra $(\g,\t)$ 
in Theorem 1, 
instead of that of the Nambu-Poisson algebra itself.


\begin{thebibliography}{99}
     \bibitem{Nam1} Y. Nambu, \PRD{7,1973,2405}.
     \bibitem{Tak1} L. Takhtajan, \CMP{160,1994,295}.
     \bibitem{Sak1} M. Sakakibara, \PTP{104,2000,1067}.
     \bibitem{Sch1} A. Schwarz, \CMP{155,1993,249}.
     \bibitem{BV1} I. Batalin and G. Vilkovisky, \PRD{28,1983,2567}.
     \bibitem{Wit1} E. Witten, Mod.\ Phys.\ Lett.\ A \textbf{7} (1990), 487.
     \bibitem{Kos1} J.-L. Koszul, Ast\'{e}risque (1985), 257.
     \bibitem{Akm1} F. Akman, q-alg/9506027. 
     \bibitem{Sta1} J. Stasheff, hep-th/9712157.
     \bibitem{Hop1} J. Hoppe, Helv. Phys. Acta. \textbf{70} (1997), 302.
\end{thebibliography}
\end{document}